\documentclass[10pt,a4paper]{IEEEtran}
\usepackage{times}
\usepackage{amsbsy}
\usepackage{latexsym}
\usepackage{amssymb}
\usepackage{amsmath}
\usepackage[usenames]{color}
\usepackage[ansinew]{inputenc}
\usepackage[dvips ]{graphicx}
\input epsf

\title{\huge Flexible Widely-Linear Multi-Branch Decision Feedback Detection Algorithms
for Massive MIMO Systems \vspace{-0.25em}}
\author{Rodrigo C.\ de Lamare \\
CETUC, Pontifical Catholic University of Rio de Janeiro, Brazil\\
Department of Electronics, University of York, York Y010 5DD, United Kingdom\\
Emails: delamare@cetuc.puc-rio.br, rcdl500@york.ac.uk}

\IEEEoverridecommandlockouts
\IEEEpubid{\makebox[\columnwidth]{978-1-4673-6540-6/15/\$31.00~\copyright~2015
IEEE \hfill} \hspace{\columnsep}\makebox[\columnwidth]{ }}

\begin{document}
\maketitle

\begin{abstract}

This paper presents widely-linear multi-branch decision feedback detection
techniques for large-scale multiuser multiple-antenna systems. We consider a
scenario with impairments in the radio-frequency chain in which the in-phase
(I) and quadrature (Q) components exhibit an imbalance, which degrades the
receiver performance and originates non-circular signals.  A widely-linear
multi-branch decision feedback receiver is developed to mitigate both the
multiuser interference and the I/Q imbalance effects. An iterative detection
and decoding scheme with the proposed receiver and convolutional codes is also
devised. Simulation results show that the proposed techniques outperform
existing algorithms.

\end{abstract}


\section{Introduction}

Large-scale multiple-antenna systems have recently attracted substantial
interest due to their potential for deployment in the next generation of
wireless networks. In particular, large-scale multiple-antenna  systems  have
the ability to offer a substantial increase in data rates and to mitigate
interference more effectively due to the extra degrees of freedom offered by
the large number of antennas \cite{marzetta_first,mmimo}. Despite the several
advantages of large-scale multiple-antenna systems, there are many problems
that need to be solved before such systems could be adopted in practice. Among
them are the development of efficient detectors that can obtain high
performance with reduced cost and operate in the presence of hardware
impairments \cite{mmimo}.

In the literature, the use of the receive matched filter (RMF) is often
advocated when the number of antennas is substantially increased and the
channels become asymptotically orthogonal \cite{mmimo,aggarwal},
\cite{mmimo,wence,Costa,delamare_ieeproc,TDS_clarke,TDS_2,switch_int,switch_mc,smce,TongW,jpais_iet,TARMO,keke1,kekecl,keke2,Tomlinson,dopeg_cl,peg_bf_iswcs,gqcpeg,peg_bf_cl,Harashima,mbthpc,zuthp,rmbthp,Hochwald,BDVP},\cite{delamare_mber,rontogiannis,delamare_itic,stspadf,choi,stbcccm,FL11,jio_mimo,peng_twc,spa,spa2,jio_mimo,P.Li,jingjing,did,bfidd,mbdf}.
Furthermore, these studies assume knowledge of the channels and seldom include
hardware impairments \cite{wence}. However, in the presence of non-orthogonal
channels the access point or the users will be affected by multiuser
interference, which requires more sophisticated detection algorithms than the
receive matched filter \cite{shepard,gao,ngo}. Linear detectors \cite{hang},
successive interference cancellation (SIC) \cite{li,peng_twc},
likelihood-ascent search (LAS) techniques \cite{vardhan,li} decision feedback
(DF) \cite{spa,jio_mimo,mbdf} and their variants are techniques that have
attractive trade-offs between performance and complexity.

With the increase of the number of antennas and associated radio frequency (RF)
chains, the size and cost-effectiveness of individual RF chains eventually
becomes critical \cite{schenk}. A technical solution to such large-scale
systems is the direct-conversion radio (DCR) architecture, which is flexible
and can operate with different air interfaces, frequency bands and waveforms.
Conversely, it does not require RF image rejection filters nor intermediate
frequency stages, resulting in lower implementation cost and smaller sizes than
the classic super-heterodyne structure. A limitation of DCR is a common RF
imperfection known as in-phase quadrature (I/Q) imbalance, which appears due to
non-ideal properties of RF mixers and leads to performance degradation.

The I/Q imbalance may also originate non-circular signals even in the presence
of circular source signals, which can be exploited by widely-linear signal
processing techniques \cite{buzzi}-\cite{hakkarainen}. Prior work on
widely-linear processing for wireless receivers includes several studies on
multiple-antenna receivers \cite{buzzi,chevalier1}, single-antenna interference
cancellation concepts \cite{chevalier2}, and large-scale multiple antenna
systems \cite{hakkarainen}.

In this work, we investigate a potential solution to large-scale
multiple-antenna systems that suffer from I/Q imbalance and exhibit
non-orthogonal channels. In particular, we combine widely-linear processing
techniques and the multi-branch concept \cite{spa,mbdf} in order to devise a
widely-linear multi-branch decision feedback (WL-MB-DF) receiver that can
mitigate I/Q imbalance and achieve a near-optimal performance. An iterative
detection and decoding scheme with the proposed receiver and convolutional
codes is also devised. Simulation results show that the proposed algorithms
outperform prior art.

The remainder of this work is structured as follows. Section II describes the
uplink of a multiuser multiple-antenna system, models the I/Q imbalance and
characterizes the second-order statistics of the received data. Section III
presents the proposed WL-MB-DF receiver. An iterative detection and decoding
scheme based on the WL-MB-DF is developed in Section IV. Section V illustrates
and discusses the simulation results, whereas Section VI gives the conclusions.

\section{System model and statistical modelling of I/Q imbalance}

In this section, we detail the uplink of a multiuser multiple-antenna system,
model the I/Q imbalance and characterize the second-order statistics of the
received data.

\subsection{Uplink system model}

We consider the uplink of a multiuser massive multiple-antenna system where the
base station employs $N_A$ antenna elements at the receiver. The scenario of
interest includes $K$ users which are equipped with $N_U$ antenna elements and
communicate with a receiver with $N_A$ antenna elements, where $N_A \geq K
N_U$. At each time instant, the $K$ users transmit $N_U$ symbols which are
organized into a $N_U \times 1$ vector ${\boldsymbol s}_k [i] = \big[
s_{k,1}[i], ~s_{k,2}[i], ~ \ldots,~ s_{k,N_U}[i] \big]^T$ taken from a
modulation constellation $A = \{ a_1,~a_2,~\ldots,~a_N \}$. The data symbols of
each user are organized in $N_U \times 1$ vectors ${\boldsymbol s}_k[i]$ and
transmitted over flat fading channels. The received signal after demodulation,
pulse matched filtering and sampling is collected in an $N_A \times 1$ vector
${\boldsymbol r}[i] = \big[ r_1[i], ~r_2[i], ~ \ldots,~ r_{N_R}[i] \big]^T$
with sufficient statistics for processing as described by
\begin{equation}
{\boldsymbol r}[i] = \sum_{k=1}^{K}{\boldsymbol H}_k {\boldsymbol s}_k[i] +
{\boldsymbol n}[i], \label{rdata}
\end{equation}
where the $N_A \times 1$ vector ${\boldsymbol n}[i]$ is a zero mean complex
circular symmetric Gaussian noise vector with covariance matrix $E\big[
{\boldsymbol n}[i] {\boldsymbol n}^H[i] \big] = \sigma_n^2 {\boldsymbol I}$.
The data vectors ${\boldsymbol s}_k[i]$ have zero mean and covariance matrices
${\boldsymbol Q}_k = E\big[ {\boldsymbol s}_k[i] {\boldsymbol s}_k^H[i] \big] =
\sigma_{s_{k}}^2 {\boldsymbol I}$, where $\sigma_{s_{k}}^2$ is the signal
power. The elements $h_{n_A,n_U}$ of the $N_A \times N_U$ channel matrices
${\boldsymbol H}_k$ represent the complex channel gains from the $n_U$th
transmit antenna to the $n_A$th receive antenna.

\subsection{Modelling and characterization of I/Q imbalance}

DCR converts the received RF signal to two real-valued baseband signals,
denoted I and Q components, respectively. The conversion is often performed
with two local oscillator signals and mixers, which have equal gains and 90
degrees phase difference. However, in practice signals and mixers have a gain
mismatch and they are not in perfect phase quadrature. For this reason, the
statistics of the received signal change and the corresponding complex signal,
even if originally circular, becomes non-circular. This effect is called I/Q
imbalance \cite{schenk}. The received signal with I/Q imbalance is described by
\begin{equation}
\begin{split}
{\boldsymbol r}_{\rm IQ}[i] & = {\boldsymbol A}_1 {\boldsymbol r}[i] +
{\boldsymbol A}_2{\boldsymbol r}^{*}[i] \\ & = \sum_{k=1}^{K}{\boldsymbol
A}_1{\boldsymbol H}_k {\boldsymbol s}_k[i] + \sum_{k=1}^{K}{\boldsymbol
A}_2{\boldsymbol H}_k^* {\boldsymbol s}_k^*[i] \\ & \quad +{\boldsymbol
A}_1{\boldsymbol n}[i]+{\boldsymbol A}_2{\boldsymbol n}^*[i], \label{rdatamis}
\end{split}
\end{equation}
where the $N_A \times N_A$ diagonal matrices ${\boldsymbol A}_j = {\rm diag}
\left( A_{j,1}~ A_{j,2} \ldots A_{j,N_A} \right)$ for $j=1,2$ contain the I/Q
imbalance components with entries given by
\begin{equation}
A_{1,i} = (1+g_i e^{-j \phi_i}), ~~~ A_{2,i} = (1- g_i e^{-j \phi_i})/2,~~ i =
1, \ldots, N_A
\end{equation}
where $g_i$ represents the relative gain mismatch and for the $i$th antenna
element and $\phi_i$ corresponds to the phase mismatch between the I- and
Q-branches. Note that in the absence of I/Q imbalance we have $g_i = 1$ and
$\phi_i=0$.

In order to characterize the I/Q imbalance, let us consider the covariance and
pseudo-covariance matrices of the data. The covariance matrix of ${\boldsymbol
r}[i]$ in (\ref{rdata}) is given by
\begin{equation}
\begin{split}
{\boldsymbol R} & \triangleq E [ {\boldsymbol r}[i] {\boldsymbol r}^H[i]] =
\sum_{k=1}^{K} {\boldsymbol H}_k {\boldsymbol Q}_k {\boldsymbol H}_k^H +
\sigma_n^2{\boldsymbol I},
\end{split}
\end{equation}
and the pseudo-covariance matrix of ${\boldsymbol r}[i]$ is defined by
\begin{equation}
\begin{split}
{\boldsymbol C} & \triangleq E [ {\boldsymbol r}[i] {\boldsymbol r}^T[i]] =
\sum_{k=1}^{K} {\boldsymbol H}_k {\boldsymbol P}_k {\boldsymbol H}_k^T,
\end{split}
\end{equation}
where the $N_U \times N_U$ pseudo-covariance matrix is given by ${\boldsymbol
P}_{k} = E[{\boldsymbol s}_k {\boldsymbol s}_k^T] = \left\{
            \begin{array}{cc}
              \sigma_{s_k}^2 {\boldsymbol I} & {\rm ~for ~ BPSK,~ ASK,~ etc,} \\
              0 & {\rm ~for ~QPSK, ~QAM, ~etc.} \\
            \end{array}
          \right. $

In order to model and statistically characterize the non-circular received
signal in (\ref{rdatamis}), we rely on the second-order statistics of the
augmented data, which yields the augmented received signal described by
\begin{equation}
{\boldsymbol r}_a[i]  \triangleq \left[ \begin{array}{c}
                         {\boldsymbol r}_{\rm IQ}[i] \\
                         {\boldsymbol r}_{\rm IQ}^*[i] \\
                       \end{array}
                     \right],
\end{equation}
and the covariance matrix of ${\boldsymbol r}_a[i]$ is given by
\begin{equation}
{\boldsymbol R}_a[i]  \triangleq E[ {\boldsymbol r}_a[i] {\boldsymbol
r}_a^H[i]] = \left[    \begin{array}{c c}
                         {\boldsymbol R}_{\rm IQ} & {\boldsymbol C}_{\rm IQ}\\
                         {\boldsymbol C}_{\rm IQ}^* & {\boldsymbol R}_{\rm IQ}^*\\
                       \end{array}
                       \right],
\end{equation}
where for arbitrary gain and phase mismatches ($g_i\neq1$ and $\phi_i \neq 0$)
the covariance matrix of ${\boldsymbol r}_{\rm IQ}[i]$ is described by
${\boldsymbol R}_{\rm IQ} \triangleq {\boldsymbol A}_1{\boldsymbol
R}{\boldsymbol A}_1^H + {\boldsymbol A}_1{\boldsymbol C} {\boldsymbol A}_2^H +
{\boldsymbol A}_2 {\boldsymbol C}^* {\boldsymbol A}_1^H + {\boldsymbol A}_2
{\boldsymbol R}^* {\boldsymbol A}_2^H $ and the pseudo-covariance of
${\boldsymbol r}_{\rm IQ}[i]$ is ${\boldsymbol C}_{\rm IQ} \triangleq
{\boldsymbol A}_1{\boldsymbol C}{\boldsymbol A}_1^T + {\boldsymbol
A}_1{\boldsymbol R} {\boldsymbol A}_2^T + {\boldsymbol A}_2 {\boldsymbol R}^*
{\boldsymbol A}_1^T + {\boldsymbol A}_2 {\boldsymbol C}^T {\boldsymbol A}_2^T$.

Several other cases of interest relating the gain and phase mismatches and the
pseudo-covariance matrix ${\boldsymbol C}$ can be used to characterize
${\boldsymbol R}_{\rm IQ}$ and ${\boldsymbol C}_{\rm IQ}$, namely:
\begin{itemize}
\item{When ${\boldsymbol C}={\boldsymbol 0}$, $g_i=1$ and
$\phi_i=0$, we have ${\boldsymbol A}_1 = {\boldsymbol I}$, ${\boldsymbol A}_2 =
{\boldsymbol 0}$, ${\boldsymbol R}_{\rm IQ} = {\boldsymbol R}$ and
${\boldsymbol C}_{\rm IQ} = {\boldsymbol 0}$.}
\item{When ${\boldsymbol C}={\boldsymbol 0}$, $g_i\neq1$ and
$\phi_i \neq 0$, we have ${\boldsymbol A}_1 \neq {\boldsymbol I}$,
${\boldsymbol A}_2 \neq {\boldsymbol 0}$, ${\boldsymbol R}_{\rm IQ} =
{\boldsymbol A}_1{\boldsymbol R}{\boldsymbol A}_2^H$ and ${\boldsymbol C}_{\rm
IQ} = {\boldsymbol A}_1{\boldsymbol R}{\boldsymbol A}_2^T + {\boldsymbol
A}_2{\boldsymbol R}^*{\boldsymbol A}_1^T$.}
\item{When ${\boldsymbol C}\neq{\boldsymbol 0}$, $g_i = 1$ and
$\phi_i = 0$, we have ${\boldsymbol A}_1 = {\boldsymbol I}$, ${\boldsymbol A}_2
= {\boldsymbol 0}$, ${\boldsymbol R}_{\rm IQ} = {\boldsymbol R}$ ${\boldsymbol
C}_{\rm IQ} = {\boldsymbol C}$.}
\end{itemize}
Note that even for ${\boldsymbol C}={\boldsymbol 0}$ due to circular
transmitted data, the I/Q imbalance can result in non circular data
${\boldsymbol C}_{\rm IQ} \neq {\boldsymbol 0}$, which can be exploited by
widely-linear processing techniques.

\section{Proposed Widely-Linear Decision Feedback Detection}

\begin{figure}[!htb]
\begin{center}
\def\epsfsize#1#2{0.9\columnwidth}
\epsfbox{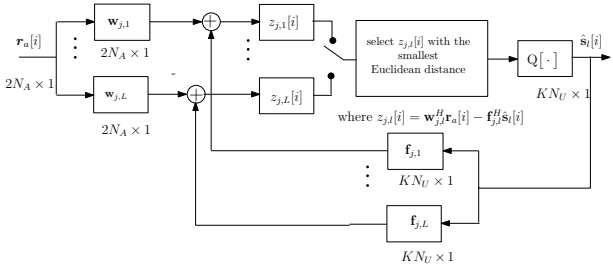} \caption{Block diagram of the proposed WL-MB-DF detector and
the processing of the $j$th data stream.} \label{mbdf}
\end{center}
\end{figure}

{In this section, the structure of the proposed WL-MB-DF detector is presented
and a schematic of the detector is shown in Fig. \ref{mbdf}. The WL-MB-DF
detector employs multiple pairs of widely-linear minimum-mean square error
(WL-MMSE) receive filters in such a way that the detector can obtain different
local maxima of the likelihood function and select the best candidate for
detection according to the Euclidean distance for each received data symbol.
The WL-MB-DF scheme is flexible and approaches the full receive diversity
available in the system by increasing the number of branches.

In order to detect each transmitted data stream using the WL-MB-DF detector,
the receiver linearly combines the feedforward filter represented by the $2N_A
\times 1$ vector ${\boldsymbol w}_{j,l}$ corresponding to the $j$-th data
stream and the $l$-th branch with ${\boldsymbol r}_a[i]$, subtracts the
remaining interference by linearly combining the feedback filter denoted by the
$KN_U \times 1$ vector ${\boldsymbol f}_{j,l}$ with the $KN_U \times 1$ vector
of initial decisions $\hat{\boldsymbol s}_l[i]$ obtained by ${\boldsymbol
w}_{j,l}$. This process is repeated for $L$ candidate symbols and $KN_U$ data
streams as described by
\begin{equation}
\begin{split}
\label{eq:df} { z}_{j,l}[i] & = {\boldsymbol w}^{H}_{j,l}{\boldsymbol r}_a[i] -
{\boldsymbol f}_{j,l}^{H} \hat{\boldsymbol s}_l[i], \\ ~~ j  & = 1, ~ \ldots,
~KN_U~~ {\rm and} ~~ l=1,~\ldots,~L,
\end{split}
\end{equation}
where the input to the decision device for the $i$th symbol and the $j$-th
stream is the $L \times 1$ vector ${\boldsymbol z}_j[i] =[z_{j,1}[i] ~\ldots
~z_{j,L}[i]]^{T}$.

The WL-MB-DF detector generates $L$ candidate symbols for each data stream and
then selects the best branch according to the Euclidean distance as described
by
\begin{equation}
\begin{split}
l_{j,{\rm opt}} & = \arg \min_{1 \leq l_j \leq L}   {\rm C} ({\boldsymbol
r}_a[i],{\boldsymbol H},\hat{\boldsymbol
s}_l[i],{\boldsymbol w}_{j,l}, {\boldsymbol f}_{j,l}), \\
~j& =1, ~\ldots ,~ KN_U, \label{eq:error}
\end{split}
\end{equation}
where
\begin{equation}{\rm C} ({\boldsymbol r}_a[i],{\boldsymbol H},\hat{\boldsymbol
s}_l[i],{\boldsymbol w}_{j,l}, {\boldsymbol f}_{j,l})) = ||{\boldsymbol r}_a[i]
- {\boldsymbol H}\hat{\boldsymbol s}_{l}[i]||
\end{equation}
is the Euclidean distance between ${\boldsymbol r}_a[i]$ and the product of the
channels of all users ${\boldsymbol H} =[{\boldsymbol H}_1 \ldots {\boldsymbol
H}_{k}]$ and the candidate symbol vector $\hat{\boldsymbol s}_l[i]$. The final
detected symbol of the WL-MB-DF detector is obtained by using the best branch:
\begin{equation}
\begin{split}
\hat{s}_{j}[i] & = Q \big[ {\boldsymbol z}_{j,l_{j,{\rm opt}}}[i] \big]
 = Q \big[ {\boldsymbol w}^{H}_{j,l_{j,{\rm opt}}}{\boldsymbol r}_a[i]
- {\boldsymbol f}_{j,l_{j,{\rm opt}}}^{H} \hat{\boldsymbol s}_{l_{j,{\rm
opt}}}[i]  \big],\\ ~j &=1, ~ \ldots,~ KN_U, \label{eq:dec}
\end{split}
\end{equation}
where $Q( \cdot)$ is a function that decides about the symbols, which can be
drawn from an M-PSK or a QAM constellation.

\subsection{WL-MMSE Filter Design }

The design of the receive filters is equivalent to determining ${\boldsymbol
w}_{j,l}$ and ${\boldsymbol f}_{j,l}$ subject to certain shape constraints on
${\boldsymbol f}_{j,l}$ in accordance to the following optimization problem
\begin{equation}
\begin{split}
\label{eq:msedfprop} {\rm min} & ~ {\rm MSE} (s_j[i],{\boldsymbol
w}_{j,l},{\boldsymbol f}_{j,l}) = E\big[ | { s}_j[i] - {\boldsymbol
w}^H_{j,l}{\boldsymbol r}_a[i] + {\boldsymbol
f}_{j,l}^{H}\hat{\boldsymbol s}_{l}[i] |^2 \big] \\
& {\rm subject}~{\rm to}~  {\boldsymbol S}_{j,l} {\boldsymbol f}_{j,l} =
{\boldsymbol 0} ~~ {\rm and} ~~
 ||{\boldsymbol f}_{j,l}||^2 = \gamma_{j,l} || {\boldsymbol
f}_{j,l}^c ||^2 , \\ {\rm for} ~ j & = 1, \ldots, KN_U ~{\rm and}~ l = 1,
\ldots, L,
\end{split}
\end{equation}
where the $KN_U \times KN_U$ shape constraint matrix is ${\boldsymbol
S}_{j,l}$, ${\boldsymbol 0}$ is a $K N_U \times 1$ constraint vector and
$\gamma_{j,l}$ is a design parameter that ranges from $0$ to $1$ and is
responsible for scaling the norm of the conventional feedback receive filter
${\boldsymbol f}_{j,l}^c$. The scaling of ${\boldsymbol f}_{j,l}^c$ results in
the desired feedback receive filter ${\boldsymbol f}_{j,l}$.

In what follows, WL-MMSE receive filters based on the proposed optimization in
(\ref{eq:msedfprop}) are derived. By resorting to the method of Lagrange
multipliers, computing the gradient vectors of the Lagrangian with respect to
${\boldsymbol w}_{j,l}$ and ${\boldsymbol f}_{j,l}$, equating them to null
vectors and rearranging the terms, we obtain ${\rm for} ~ j = 1, \ldots, KN_U$
and $l = 1, \ldots, L$
\begin{equation}
\label{eq:dfeprop1} {\boldsymbol w}_{j,l}^{\rm MMSE} = {\boldsymbol
R}_a^{-1}({\boldsymbol p}_{a,j} + {\boldsymbol Q}_a{\boldsymbol f}_{j,l}),
\end{equation}  {
\begin{equation}
\begin{split}
\label{eq:dfeprop2} {\boldsymbol f}_{j,l}^{\rm MMSE} & =
\frac{\beta_{j,l}}{\sigma_s^2} {\boldsymbol \Pi}_{j,l} {\boldsymbol
Q}^H_a{\boldsymbol w}_{j,l},
\end{split}
\end{equation} }
where the $2N_A \times 1$ augmented cross-correlation vector is
\begin{equation}
\begin{split}
{\boldsymbol p}_{a,j} \triangleq E[{\boldsymbol r}_a[i] s_j^*[i]] = \left(
\begin{array}{c}
          {\boldsymbol A}_1 {\boldsymbol H}_j {\boldsymbol q}_j + {\boldsymbol A}_2{\boldsymbol H}_j^*{\boldsymbol p}_j \\
          {\boldsymbol A}_1^* {\boldsymbol H}_j^* {\boldsymbol p}_j + {\boldsymbol A}_2^*{\boldsymbol H}_j{\boldsymbol q}_j \\
        \end{array}
      \right),
\end{split}
\end{equation}
where ${\boldsymbol q}_j = E[{\boldsymbol s}_k[i] s_j^*[i]] = \sigma_{s_k}^2
{\boldsymbol t}_j$, ${\boldsymbol t}_j$ is an $N_U \times 1$ vector with a one
in the $j$th entry and zeros elsewhere, the $N_U \times 1$ cross-correlation
vector is $${\boldsymbol p}_j = E[{\boldsymbol s}_k^*[i]s_j^*] = \left\{
                       \begin{array}{cc}
                         \sigma_{s_k}^2 {\boldsymbol t}_j & {\rm ~for ~BPSK, ~ASK,~etc}, \\
                         {\boldsymbol 0} ~({\boldsymbol C}={\boldsymbol 0}) &  {\rm ~for~ QPSK, ~QAM, ~etc} \\
                       \end{array}
                     \right.$$ and the $ 2N_A \times K N_U$
cross-correlation matrix is given by
\begin{equation}
{\boldsymbol Q}_a \triangleq E\big[ {\boldsymbol r}_a[i] \hat{\boldsymbol
s}_{l}^{H}[i] \big] = \left(
\begin{array}{c}
          \sigma_{s_k}^2{\boldsymbol A}_1 {\boldsymbol H} + \sigma_{s_k}^2{\boldsymbol A}_2{\boldsymbol H}^* \\
          \sigma_{s_k}^2{\boldsymbol A}_1^* {\boldsymbol H}_j^* + \sigma_{s_k}^2{\boldsymbol A}_2^*{\boldsymbol H} \\
        \end{array}
      \right),
\end{equation}
where
\begin{equation}
{\boldsymbol \Pi}_{j,l} = {\boldsymbol I} - {\boldsymbol
S}_{j,l}^H({\boldsymbol S}_{j,l}^H {\boldsymbol S}_{j,l})^{-1}{\boldsymbol
S}_{j,l}
\end{equation}
{is a projection matrix that ensures the shape constraint ${\boldsymbol
S}_{j,l}$ on the feedback filter, $\beta_{j,l} = (1 - \mu_{j,l})^{-1}$ is the
parameter that controls the ability of the WL-MB-DF detector to mitigate error
propagation with values $0 \leq \beta_{j,l} \leq 1$, and $\mu_{j,l}$ is the
Lagrange multiplier. Note that the inverse $({\boldsymbol S}_{j,l}^H
{\boldsymbol S}_{j,l})^{-1}$ might not exist. In these situations, a
pseudo-inverse is computed. The relationship between $\beta_{j,l}$ and
$\gamma_{j,l}$ is not in closed-form except for the extreme values when we have
$\beta_{j,l} =0$ and $\beta_{j,l} =1 $ for $\gamma_{j,l}=0$ (standard WL MMSE
detector) and $\gamma_{j,l}=1$ (standard WL-MB-DF detector), respectively. The
above expressions only depend on statistical quantities, and consequently on
the channel matrix ${\boldsymbol H}$, the symbol and noise variances
$\sigma^2_{s_j}$ and $\sigma^2_n$, respectively, and the constraints.

The MMSE associated with the filters ${\boldsymbol w}_{j,l}^{\rm MMSE}$ and
${\boldsymbol f}_{j,l}^{\rm MMSE}$ and the statistics of the data symbols
$s_j[i]$ is given by
\begin{equation}
\begin{split}
\underbrace{{\rm MMSE} (s_j[i],{\boldsymbol w}_{j,l}^{\rm MMSE}, {\boldsymbol
f}_{j,l}^{\rm MMSE})}_{{\rm MMSE}_j}  & = \sigma_{s_j}^2 - {\boldsymbol
w}_{j,l}^{H, ~{\rm MMSE}} {\boldsymbol R}_a {\boldsymbol w}_{j,l}^{\rm MMSE}
\\ & \quad + {\boldsymbol f}_{j,l}^{H, ~{\rm MMSE}}
{\boldsymbol f}_{j,l}^{\rm MMSE}, 
\end{split}
\end{equation}
where $\sigma_{s_j}^2 = E[|s_j[i]|^2]$ is the variance of the desired symbol.
The receive filters can be computed with more sophisticated algorithms
\cite{scharf,bar-ness,pados99,reed98,hua,goldstein,santos,qian,delamarespl07,xutsa,delamaretsp,kwak,xu&liu,delamareccm,wcccm,delamareelb,jidf,delamarecl,delamaresp,delamaretvt,jioel,delamarespl07,delamare_ccmmswf,jidf_echo,delamaretvt10,delamaretvt2011ST,delamare10,fa10,lei09,ccmavf,lei10,jio_ccm,ccmavf,stap_jio,zhaocheng,zhaocheng2,arh_eusipco,arh_taes,dfjio,rdrab,dcg_conf,dcg,dce,drr_conf,dta_conf1,dta_conf2,dta_ls,song,wljio,barc,jiomber,saalt}.

\subsection{Design of Shape Constraint Matrices and Ordering}

The shape constraint matrices ${\boldsymbol S}_{j,l}$ modify the structure of
the feedback filters ${\boldsymbol f}_{j,l}$ in such a way that only the
selected feedback elements of ${\boldsymbol f}_{j,l}$ will be used to cancel
the interference between the data streams. The feedback connections perform
interference cancellation with a chosen ordering. For the first branch of
detection ($l=1$), we employ
\begin{equation}
\begin{split}
{\boldsymbol S}_{j,l} {\boldsymbol f}_{j,l} & = {\boldsymbol 0}, ~~
l=1, ~ j=1, \ldots, KN_U,\\
{\boldsymbol S}_{j,l} & = \left[ \begin{array}{cc} {\boldsymbol
0}_{KN_{U}-j+1,KN_{U}-j+1} & {\boldsymbol 0}_{ KN_{U}-j+1, j-1} \\
{\boldsymbol 0}_{j-1, KN_{U}-j+1} & {\boldsymbol I}_{j-1,j-1}
\end{array} \right],
\end{split}
\end{equation}
where ${\boldsymbol 0}_{m,n}$ denotes an $m \times n$-dimensional matrix full
of zeros, and ${\boldsymbol I}_m$ denotes an $m$-dimensional identity matrix.
For the remaining branches, an approach based on permutations in the matrices
${\boldsymbol S}_{j,l}$ is adopted, which is given by
\begin{equation}
\begin{split}
{\boldsymbol S}_{j,l} {\boldsymbol f}_{j,l} & = {\boldsymbol 0}, ~~ l=2,
\ldots, L, ~
j=1, \ldots, KN_U,\\
{\boldsymbol S}_{j,l} & = \phi_l \left[ \begin{array}{cc} {\boldsymbol
0}_{KN_{U}-j+1,KN_{U}-j+1} & {\boldsymbol 0}_{ KN_{U}-j+1, j-1} \\
{\boldsymbol 0}_{j-1, KN_{U}-j+1} & {\boldsymbol I}_{j-1,j-1}
\end{array} \right],
\end{split}
\end{equation}
where the operator $\phi_l[ \cdot ]$ permutes the elements of the argument
matrix such that this results in different cancellation patterns.

For the first branch, an ordering algorithm based on increasing values of the
MMSE is considered. The ordering of the remaining branches depends on the
maximization of the difference between the MMSE of different data streams:
\begin{equation}
\begin{split}
o_{j,l} & = \arg \max_{n}  \sum_{q=1}^{j-1}|{\rm MMSE}_{n} - {\rm
MMSE}_{o_{j,q}} |, \\ {\rm for} ~ l & = 2,~ \ldots, ~L ~{\rm and}~ j,n=1,
\ldots, KN_U \\ ~~ & {\rm subject ~to} ~~  {\rm MMSE}_{o_{j,l}} \neq {\rm
MMSE}_{o_{q,l}},~ q =1,\ldots, j-1. \label{sub_ord}
\end{split}
\end{equation}}

\section{Iterative Processing}

This section presents an iterative version of the proposed WL-MB-DF detector
operating with soft-input soft-output detection and decoding, and with
convolutional codes \cite{choi}-\cite{wang}. The receiver structure consists of
the following stages: a soft-input-soft-output (SISO) WL-MB-DF detector and a
maximum \textit{a posteriori} (MAP) decoder. These stages are separated by
interleavers and deinterleavers. The soft outputs from the WL-MB-DF are used to
estimate log-likelihood ratios (LLRs) which are interleaved and serve as input
to the MAP decoder for the convolutional code. The MAP decoder computes
\textit{a posteriori} probabilities (APPs) for each stream's encoded symbols,
which are used to generate soft estimates. These soft estimates are used to
update the receive filters of the WL-MB-DF detector, de-interleaved and fed
back through the feedback filter. The WL-MB-DF detector computes the \textit{a
posteriori} log-likelihood ratio (LLR) of a transmitted symbol ($+1$ or $-1$)
for every code bit of each stream as given by
\begin{equation}
\begin{split}
\Lambda_1[b_{j,c,l}[i]] & = {\rm log} \frac{P[{\boldsymbol
r}_a[i]|b_{j,c,l}[i]=+1]}{P[ {\boldsymbol r}_a[i]|b_{j,c,l}[i]=-1]} + {\rm log}
\frac{P[b_{j,c}[i]=+1]}{P[b_{j,c}[i]=-1]} \\  & = \lambda_1[b_{j,c,l}[i]] +
\lambda_2^p[b_{j,c}[i]], \\  j &= 1, \ldots, KN_U, ~c  =1, \ldots,  C,  ~l=1,
\ldots,L, \label{llr}
\end{split}
\end{equation}
where $C$ is the number of bits used to map the constellation,
$\lambda_2^p[b_{j,c}[i]] = {\rm log} \frac{P[b_{j,c}[i]=+1]}{P[b_{j,c}[i]=-1]}$
is the \textit{a priori} LLR of the code bit $b_{j,c}[i]$, which is computed by
the MAP decoder processing the $j$th data stream in the previous iteration,
interleaved and then fed back to the WL-MB-DF detector. The superscript $^p$
denotes the quantity obtained in the previous iteration. Assuming equally
likely bits, we have $\lambda_2^p[b_{j,c}[i]] =0$ in the first iteration for
all streams. The quantity $\lambda_1[b_{j,c,l}[i]] = {\rm log}
\frac{P[{\boldsymbol r}_{a}[i]|b_{j,c,l}[i]=+1]}{P[ {\boldsymbol
r}_a[i]|b_{j,c,l}[i]=-1]}$ represents the \textit{extrinsic} information
computed by the detector based on the received data ${\bf r}_a[i]$, and the
prior information about the code bits $\lambda_2^p[b_{j,c}[i]],~
j=1,\ldots, KN_U,~ c=1,\ldots, C$  and the $i$th data symbol. 

For the MAP decoding, we assume that the interference plus noise at the output
${\boldsymbol z}_{j,l}[i]$ of the linear receive filters is Gaussian. Thus, for
the $j$th stream, the $l$th branch and the $q$th iteration the soft output of
the WL-MB-DF detector is
\begin{equation}
z_{j,l}^{(q)}[i] = V_{j,l}^{(q)} s_{j,l}[i] + \xi_{j,l}^{(q)}[i],
\label{output}
\end{equation}
where $V_{j,l}^{(q)}[i]$ is a scalar variable equivalent to the magnitude of
the channel corresponding to the $j$th data stream and $\xi_{j,l}^{(q)}[i]$ is
a Gaussian random variable with variance $\sigma^2_{\xi_{j,l}^{(q)}}$. Since we
have
$V_{j,l}^{(q)}[i] = E\big[ s_{j,l}^*[i] z_{j,l}^{(q)}[i] \big]$
and
$\sigma^2_{\xi_{j,l}^{(q)}}[i]  = E\big[ | z_{j,l}^{(q)}[i] - V_{j,l}^{(q)}[i]
s_{j,l}[i]|^2 \big]$,
the receiver can obtain the estimates ${\hat V}_{j,l}^{(q)}[i]$ and ${\hat
\sigma}^2_{\xi_{j,l}^{(q)}}[i]$ via time averages. These estimates are used to
compute the \textit{a posteriori} probabilities $P[b_{j,c,l}[i] = \pm 1 |
z_{j,l}^{(q)}[i]]$ which are de-interleaved and used as input to the MAP
decoder. In what follows, it is assumed that the MAP decoder generates APPs
$P[b_{j,c,l}[i] = \pm 1]$, which are used to compute the input to the feedback
filter ${\boldsymbol f}_{j,l}$. From (\ref{output}) the extrinsic information
generated by the iterative WL-MB-DF is given by
\begin{equation}
\begin{split}
\lambda_1[b_{j,c,l}[i]] & 
= \log \frac{\sum\limits_{{\mathbb S} \in {\mathbb S}_c^{+1}} \exp \Big(
-\frac{|z_{j,l}^{(q)}[i] - V_{j,l}^{(q)}{\mathbb S}|^2}{ 2
\sigma^2_{\xi_{j,l}^{(q)}}[i]} \Big)} {\sum\limits_{{\mathbb S} \in {\mathbb
S}_c^{-1}} \exp \Big( -\frac{|z_{j,l}^{(q)}[i] - V_{j,l}^{(q)}{\mathbb S}|^2}{
2 \sigma^2_{\xi_{j,l}^{(q)}}[i]} \Big)},
\end{split}
\end{equation}
where ${\mathbb S}_c^{+1}$ and ${\mathbb S}_c^{-1}$ are the sets of all
possible constellations that a symbol can take on such that the $c$th bit is
$1$ and $-1$, respectively. The iterative WL-MB-DF detector chooses the LLR
from a list of $L$ candidates for the decoding iteration as}
\begin{equation}
\lambda_1[b_{j,c,l_{\rm opt}}[i]]  = \arg \max_{1 \leq l \leq L}
\lambda_1[b_{j,c,l}[i]],
\end{equation}
where the selected estimate is the value $\lambda_1[b_{j,c,l_{\rm opt}}[i]]$
which maximizes the likelihood and corresponds to the most likely bit. Based on
the selected prior information $\lambda_1^p[b_{j,c,l_{\rm opt}}[i]]$ and the
trellis structure of the code, the MAP decoder processing the $j$th data stream
and the $l$th branch computes the \textit{a posteriori} LLR of each coded bit
as described by
\begin{equation}
\begin{split}
\Lambda_2[b_{j,c}[i]] & = {\rm log} \frac{P[b_{j,c}[i]=+1|
\lambda_1^p[b_{j,c,l_{\rm opt}}[i]; {\rm decoding}]}{P[b_{j,c}[i]=-1|
\lambda_1^p[b_{j,c,l_{\rm opt}}[i]; {\rm decoding}]}
\\ & = \lambda_2[b_{j,c}[i]] + \lambda_1^p[b_{j,c,l_{\rm opt}}[i]]. 
\end{split}
\end{equation}
Note that the output of the MAP decoder is the sum of the prior information
$\lambda_1^p[b_{j,c,l_{\rm opt}}[i]]$ and the extrinsic information
$\lambda_2[b_{j,c}[i]]$ produced by the MAP decoder. This extrinsic information
is the information about the coded bit $b_{j,c}[i]$ obtained from the selected
prior information about the other coded bits $\lambda_1^p[b_{j,c,l_{\rm
opt}}[k]], ~ j \neq i$ \cite{wang}. The MAP decoder also computes the \textit{a
posteriori} LLR of every information bit, which is used to make a decision on
the decoded bit at the last iteration. After interleaving, the extrinsic
information obtained by the MAP decoder $\lambda_2[b_{j,c}[i]]$ for $j=1,
\ldots KN_U$, $c=1, \dots, C$ is fed back to the WL-MB-DF detector, as the
prior information about the coded bits of all streams in the subsequent
iteration.

\section{Simulations}

In this section, the bit error ratio~(BER) performance of the WL-MB-DF and
other relevant MIMO detection schemes is evaluated. The matched-filter, the SIC
receivers \cite{choi} with linear and WL-MMSE receive filters, the LAS
algorithm \cite{vardhan,li} with linear MMSE receive filters, the MB-MMSE-DF
\cite{mbdf} and the proposed WL-MB-DF techniques with error propagation
mitigation techniques are considered in the simulations. The channel
coefficients are static and obtained from complex Gaussian random variables
with zero mean and unit variance. The modulation employed is either QPSK. 
I/Q imbalance in the RF chains was implemented as a random unequal I/Q
imbalance in receiver branches ($g_i$ and $\phi_i$ were uniformly distributed
in $[0.85, 1.15]$ and $[-15^0, 15^0]$, respectively). It is assumed that all
parallel receiver branches have their own hardware. Both uncoded and coded
systems are considered. For the coded systems and iterative detection and
decoding, a non-recursive convolutional code with rate $R=1/2$, constraint
length $3$, generator polynomial $g = [ 7~ 5 ]_{\rm oct}$ and $5$ decoding
iterations is adopted. The numerical results are averaged over $10^6$ runs,
packets with $Q=500$ symbols for uncoded systems and $Q=1000$ coded symbols are
employed and the signal-to-noise ratio (SNR) in dB is defined as $\textrm{SNR}
= 10 \log_{10} \frac{KN_U \sigma_{s_k}^2}{R
 C ~\sigma^2}$, where $R<1$ is the rate of the channel
code and $C$ is the number of bits used to represent the constellation.

\begin{figure}[!htb]
\begin{center}
\def\epsfsize#1#2{0.9\columnwidth}
\epsfbox{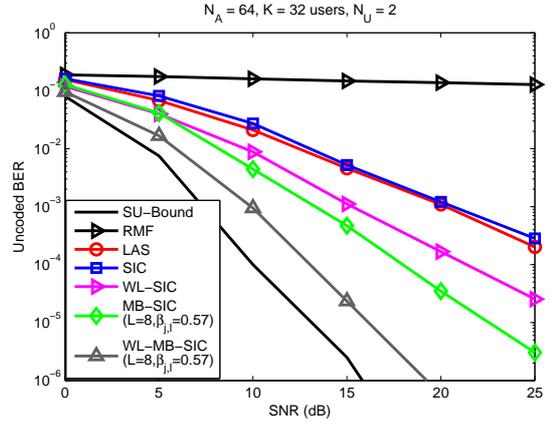} \vspace{-0.5em}\caption{Uncoded BER performance of the
proposed WL-MB-DF detector with QPSK and the presence of I/Q imbalance.}
\label{uber}
\end{center}
\end{figure}

The uncoded BER performance of the proposed WL-MB-DF detector and existing
schemes is considered with $L=8$ branches and optimized $\beta_{j,l}$. The
results shown in Fig. \ref{uber} indicate that WL-MB-DF outperforms MB-DF by up
to 5dB in terms of required SNR for the same BER performance, which is followed
by the WL-SIC, LAS, SIC, and the RMF detectors.

\begin{figure}[!htb]
\begin{center}
\def\epsfsize#1#2{0.9\columnwidth}
\epsfbox{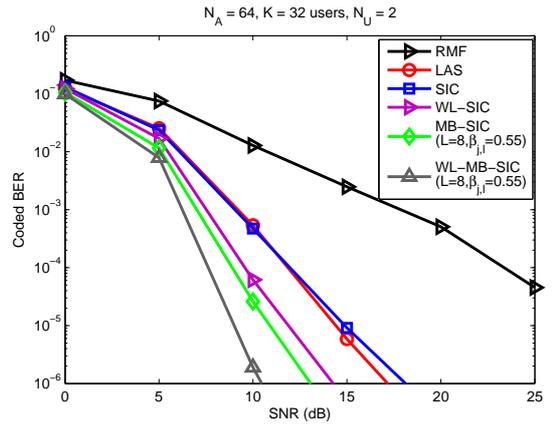} \vspace{-0.5em}\caption{Coded BER performance of the
proposed WL-MB-DF detector with QPSK and the presence of I/Q imbalance.}
\label{cber}
\end{center}
\end{figure}

The coded BER performance of the proposed WL-MB-DF detector and existing
schemes is then considered with $L=8$ branches and optimized $\beta_{j,l}$. The
results shown in Fig. \ref{cber} indicate that the same performance hierarchy
is observed and that I/Q imbalance is responsible for a significant performance
degradation of the detector with standard linear MMSE filters.

\section{Conclusions}

We have proposed and studied the WL-MB-DF detector for large-scale
multiple-antenna systems in the presence of I/Q imbalance. The results have
shown that WL-MB-DF can achieve a near-ML performance and mitigate the I/Q
imbalance, resulting in significant performance gains over prior art.

\end{document}